\documentclass[twocolumn,preprintnumbers,nofootinbib]{revtex4}
\usepackage{graphicx}
\usepackage{amssymb}
\usepackage{color}
\def\beq{\begin{equation}}
\def\eeq{\end{equation}}
\def\beqn{\begin{eqnarray}}
\def\eeqn{\end{eqnarray}}

\begin {document}
\title{Particle Creation by Loop Black Holes}
\author{Emanuele Alesci}
\affiliation{Universit\"at Erlangen, Institut f\"ur Theoretische Physik III,
 Lehrstuhl f\"ur Quantengravitation, Staudtstrasse 7, D-91058 Erlangen, EU }
 \author{Leonardo Modesto}
 \affiliation{Perimeter Institute for Theoretical Physics, 31 Caroline St.N., Waterloo, ON N2L 2Y5, Canada 
}


\begin{abstract}
We study the black hole particle production in a regular spacetime metric obtained in a 
minisuperspace approach to loop quantum gravity. In different 
previous papers the static solution was obtained and shown to
be singularity-free and self-dual. 
In this paper expanding a previous study 
of the black hole dynamics
we repeat the Hawking analysis 
which leads to a thermal flux of particles at the future infinity.
The evaporation time is infinite and the unitarity is recovered 
 due to the regularity of the 
 spacetime and to the characteristic behavior of the surface gravity.

\end{abstract}
\pacs{04.70.-s, 04.20.Dw}
\maketitle

\section{Introduction}

Black holes are one of the most fascinating predictions of Einsten's gravitational 
theory. 
Today we know that they are not just a mathematically
possible solution of general relativity, but part of Nature.
Since more than a decade now, we have good evidence that our Milky Way, as other
galaxies, hosts many stellar black holes as well as a supermassive black hole in its
center.

From the perspective of quantum gravity, black holes are of interest because of
the infinite curvature towards their center which signals (probably) a
breakdown of General Relativity. It is an area where quantum gravity effects
are strong, and it is generally expected that  these
prevent the formation of the singularity. 

In the mid-seventies Steven Hawking \cite{Haw1} and Jacob Bekenstein \cite{beke} showed that since classical
black holes radiate particles, their mass slowly decrease and finally disappear, leaving a paradox.
Supposing that matter which enters the black hole was in a pure quantum state at early time, then the existence of an horizon transforms such matter into the mix thermal state of Hawking radiation at late time for an external observer, and the complete evaporation of the singularity destroys the correlations between the radiation and the information ``swallowed" by the black hole that would allow to reconstruct the original pure quantum state.
The role of the singularity is crucial in this paradox. Suppose there is an entangled pure state
at early time, and a part of the entangled system ends inside the black hole while the other 
part remains outside the event horizon, the result is a mixed state after the partial trace is taken 
over the interior of the black hole. Now, since everything that falls inside the black hole reaches the 
singularity ($r=0$) in finite proper time, the part of the Hilbert space that is traced over 
disappears and never appears again with the disappearence of the black hole. 
In the standard semiclassical treatment this is the scenario,
since the black hole emits
particles in the form of Hawking radiation and the horizon
radius decreases and approaches the singularity
until both, singularity and horizon, vanish in the
endpoint of evaporation \cite{Haw2}. However, if the singularity does not
exist, this scenario cannot be correct. Since the singularity plays
a central role for the causal spacetime diagram, its absence in the
presence of quantum gravitational effects has consequences for the
entire global structure \cite{AB}, and its removal is
essential for resolving the black hole information loss problem \cite{Infloss,2d}. To understand the dynamics of
the gravitational and matter fields, it is then necessary to have a concrete model.

It is thus promising that a
resolution of the big bang as well as the black
hole singularites \cite{Ashtekar:2005qt,M1,M2} has been achieved in a simplified version
of loop quantum gravity (LQG)  \cite{lqgcan}, known as loop quantum cosmology (LQC) \cite{Bojowald:2006da} . The regular static
black hole metric was recently derived in \cite{Modesto:2008im}\cite{Mode3}, and
studied more closely in \cite{Modesto:2009ve}.  
The polymeric approach to solve the black hole singularity problem was also 
applied in \cite{poly}. 
An alternative resolution of the black hole 
singularity was obtained in an effective, noncommutative approach to quantum gravity 
\cite{Nicolini:2005vd} and in asymptotically safe quantum gravity \cite{RB}.
In others works \cite{2d}\cite{poly2}, a
2-dimensional model was used to study the evaporation process in the absence
of a singularity. Recently also regular spinning loop black holes were 
obtained in \cite{SLBH}. 

Here, we will use a 4-dimensional model based on the static solution
derived in \cite{Modesto:2008im} and generalize it to a dynamical case
which then allows us to examine the causal structure. This generalization
holds to good accuracy in all realistic scenarios. This approach
should be understood not as an exact solution to a problem that requires
knowledge of a full theory of quantum gravity, but as a plausible model 
based on preliminary studies that allows us to investigate the
general features of such regular black hole solutions  with quantum gravity corrections inspired by LQG.
The main goal of this paper is to reproduce the Hawking calculation of particle 
creation in a particular regular black hole background and to show that 
the whole process, collapse and complete evaporation, is unitary.
In this analysis we will make extensively use of the Fabbri and Navarro-Salas book
\cite{Fabbri}.

Non-singular black holes were considered already by
Bardeen in the late 60s and have a long history \cite{Bar,Frolov:1989pf,Frolov:1988vj,Balbinot:1990zz,Aurilia:1990uq,Bar2,Bar3,Bar4,Bar5,Bar6,Hayward:2005gi,Nicolini:2005zi,Ansoldi:2006vg,Spallucci:2008ez,Nicolini:2009gw,Nicolini:2008aj}. In this paper we extend the 
procedure in \cite{Hayward:2005gi} \cite{SIM} analyzing the particle 
production and the unitary problem as result of the complete evaporation.
The paper is organized as follows. We start in the section \ref{static}
by recalling the regular static metric we will use in the whole article. In section
\ref{dynamical} we generalize it to a collapse scenario as already done in \cite{SIM}.
In section \ref{scalarfield} we derive and solve the scalar field equations of motion 
in this regular black hole background and 
in section \ref{creation} we use these solutions to reproduce the Hawking analysis leading to a thermal spectrum. 
In section \ref{evap} we summarize the complete dynamics for the LBH collapse and evaporation. We dedicate section \ref{unitarity problem} to review the basic aspects of the unitarity and information loss problems associated to classical black holes 
 and finally in section  \ref{unitary}  we show how in this quantum corrected model these problems are solved.
The signature of the metric used in this paper is $(-,+,+,+)$ and we use the natural unit
convention $\hbar=c=G_N=1$.

\section{The Regular Black Hole spacetime}
\label{static}

LQG is a candidate theory of quantum gravity. It is obtained from the canonical quantization of the Einstein equations written in terms of the Ashtekar variables \cite{AA}, that is in terms of an $\mathrm{su}(2)$ 3-dimensional connection $A$ and a triad $E$. The result \cite{Mercuri:2010xz} is that the basis states of LQG are closed graphs made of edges and nodes labelled by irreducible $\mathrm{SU}(2)$ representations and intertwiners respectively. Physically, the edges represent quanta of area with area $8 \pi \gamma l_{\rm P}^2 \sqrt{j(j+1)}$, where $j$ is the representation label of the edge (an half-integer), $l_{\rm P}$ is the Planck length,  and $\gamma$ is a parameter of order 1 called the Immirzi parameter. Vertices of the graph represent quanta of 3-volume. 
The area is quantized and the smallest possible quanta correspond to an area 
\begin{equation}
A_{\mathrm{min}} = 4 \pi \sqrt{3} \gamma l_P^2 \, .
\end{equation}

To obtain the simplified black hole model, the following assumptions were made. First, the number of variables was reduced by assuming spherical symmetry. Second, instead of all possible closed graphs, a regular lattice with edge-lengths $ \approx \delta_b$ and $\delta_c$ was used. The dynamical solution inside the homogeneous region (that is, inside the horizon, where space is homogeneous but not static) was then obtained. An analytic continuation to the region outside the horizon shows that one can reduce the two free parameters by identifying the minimum area in the solution with the minimum area of LQG.
The remaining unknown constant of the model, $\delta_b$,  is the dimensionless polymeric
parameter. This determines, with $A_{\rm min}$, the strength of deviations from the classical theory and must be constrained by experiment.

The procedure to obtain the metric is in short the following.
\begin{enumerate}

\item We define the Hamiltonian constraint replacing the homogeneous connection with 
the holonomies along the fixed graph decided above. The diff-constraint is then identically zero 
because of homogeneity and the Gauss constraint is zero for the Kantowski-Sachs spacetime. 
\item We solve the Hamilton equation of motion for the holonomic Hamiltonian system
imposing the Hamiltonian constraint to be zero. 
\item The third step consists to extend the solution to the whole spacetime; this step is mathematically 
correct but, since we found the solution in the homogeneous region, we can not
be sure of the correct Hamiltonian constraint polymerization in the full spacetime.
However we believe such polymerization exists. 
\end{enumerate}

This quantum gravitationally corrected
Schwarzschild metric can be expressed in the form
\begin{eqnarray}
&& ds^2 = - G(r) dt^2 + \frac{dr^2}{F(r)} + H(r) d\Omega^2~, \nonumber \\
&& G(r) = \frac{(r-r_+)(r-r_-)(r+ r_{*})^2}{r^4 +a_o^2}~ , \nonumber \\
&& F(r) = \frac{(r-r_+)(r-r_-) r^4}{(r+ r_{*})^2 (r^4 +a_o^2)} ~, \nonumber \\
&& H(r) = r^2 + \frac{a_o^2}{r^2}~ .
\label{g}
\end{eqnarray}
with $d \Omega^2 = d \theta^2 + \sin^2 \theta d \phi^2$.
Here, $r_+ = 2m$ and $r_-= 2 m P^2$ are the two horizons, and $r_* = \sqrt{r_+ r_-} = 2mP$. $P$ is the
polymeric function $P = (\sqrt{1+\epsilon^2} -1)/(\sqrt{1+\epsilon^2} +1)$, with
$\epsilon \ll 1$ being the product of the Immirzi parameter ($\gamma$) and the polymeric parameter ($\delta$). With this, it is 
also $P \ll 1$, such that $r_-$ and $r_*$ are very close to $r=0$. The area $a_o$ is equal to $A_{\rm min}/8 \pi$, $A_{\rm min}$ being the minimum area gap of LQG.
Note that in the above metric, $r$ is only asymptotically the usual radial
coordinate since $H(r)$ is not just $r^2$. This choice of
coordinates however has the advantage of easily revealing the properties
of this metric as we will see. 
The ADM mass is the mass inferred by an observer at flat asymptotic infinity; it is determined solely 
by the metric at asymptotic infinity.  The parameter $m$ in the solution is related to the mass $M$ by $M = m (1+P)^2$.

If one now makes the coordinate transformation $R = a_o/r$ with the rescaling 
$\tilde t= t \, r_*^{2}/a_o$, and
simultaneously substitutes $R_\pm = a_o/r_\mp$, $R_* = a_o/r_*$ one finds that the metric in
the new coordinates has the same form as in the old coordinates and thus exhibits a
very compelling type of self-duality with dual radius $r=\sqrt{a_o}$. Looking at the angular part
of the metric, one sees that this dual radius corresponds to a minimal possible 
surface element. It is then also clear that in the limit $r\to 0$, corresponding
to $R\to \infty$, the solution
does not have a singularity, but instead has another asymptotically flat Schwarzschild region.
The dual mass observed from the dual observer in $r \approx 0$ is $m_d = a_o (1+P)^2/4 m P^2$.

\begin{figure}
\includegraphics[width=6cm]{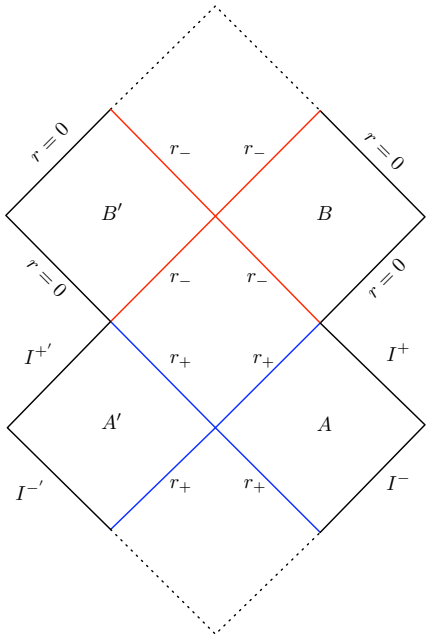}
\caption{Penrose diagram of the regular static black hole solution with two
asymptotically flat regions. Both horizons, located at $r_+$ and $r_-$,
are marked in blue and red respectively. }
\label{bhh0}
\end{figure}

The causal diagram for this metric, shown in Fig \ref{bhh0}, then has two horizons and two pairs of asymptotically
flat regions, $A, A'$ and $B,B'$, as opposed to one such pair in the standard case. In the region enclosed by 
the horizons, space- and timelikeness
is interchanged. The horizon at $r_+$ is a future horizon for observers in the asymptotically flat $A,A'$ region  and a past horizon for observers inside the two horizons. Similarly, the $r_-$ horizon is a future horizon for observers inside the two horizons but a past horizon for observes in $B, B'$.  If one computes the time
it takes for a particle to reach $r=0$, one finds that it takes infinitely long \cite{Modesto:2009ve}. The diagram shown in Fig.\ref{bhh0} 
can be analytically continued on the dotted horizons at the bottom and top. 


The metric in Eq. (\ref{g}) is a solution of a quantum gravitationally corrected set of
equations that, in the $\epsilon, a_o \to 0$ limit, reproduce Einstein's field equations.
However, due to these quantum corrections, the
above metric is no longer a vacuum-solution to Einstein's field
equations. Instead, if one computes the Einstein-tensor and sets it
equal to a source term $G_{\mu \nu} = 8 \pi \widetilde{T}_{\mu \nu}$, one
obtains an effective quantum gravitational stress-energy-tensor $\widetilde{T}_{\mu \nu}$.
The exact expressions for the components of $\widetilde{T}$ are
somewhat unsightly and can
be found in the appendix of \cite{SIM}. 
For our purposes it is here
sufficient to note that the entries are not positive definite and
violate the positive energy condition which is one of the assumptions
for the singularity theorems.

We can write the metric (\ref{g}) introducing the physical radius as radial coordinate.
This is possible introducing the new radial coordinate $x = \sqrt{H}$ which varies in the range
 $x \in [ \sqrt{2 a_o}, + \infty[$
\begin{eqnarray} 
x = \sqrt{r^2 +\frac{a_o^2}{r^2}} \, , \,\,\,\,  x \in [ \sqrt{2 a_o}, + \infty[.  
\label{x}
\end{eqnarray}
The metric assumes the following form
\begin{eqnarray}
ds^2 = - G(r(x)) dt^2 + \frac{1}{F(r(x))} \left(\frac{d r }{d x} \right)^2 dx^2 + x^2 d \Omega^2,
\label{metricmass}
\end{eqnarray}
where $G$, $F$ are implicit functions of $x$.
It is usual to define $g_{xx}$ to be related to the mass inside a sphere of radius $x$ and 
the $g_{tt}$ component of the metric to be proportional to a {\em dirty} factor $e^{2 \Phi(x)}$,
since 
$g_{tt} \neq - 1/g_{xx}$,
\begin{eqnarray}
&& g_{xx} = \left(1 - \frac{2 m(x)}{x} \right)^{-1} \nonumber, \\
&& g_{tt} = - G =  - \left(1 - \frac{2 m(x)}{x} \right) e^{2 \Phi(x)}
\label{gxx}
\end{eqnarray}
and the {\em dirty} factor is 
\begin{eqnarray}
e^{2 \Phi(x)} = \frac{(r(x) + r_*)^4 (r^4(x) + a_o^2)}{(r^4(x) - a_o^2)^2}.
\label{dirty}
\end{eqnarray}
In the radial $x$ coordinate the self-dual metric reads 
\begin{eqnarray}
ds^2 = - e^{2 \Phi(x)} \left(1 - \frac{2 m(x)}{x} \right)  dt^2 + \frac{dx^2}{1 - \frac{2 m(x)}{x}} + x^2 d \Omega^2.
\nonumber 
\end{eqnarray}
The metric in this form  satisfies the following Einstein's equations 
\begin{eqnarray}
&& \frac{ d m}{dx} = 4 \pi \rho \, x^2 ,  \nonumber \\
&& \frac{1}{G} \frac{ d G}{dx} = \frac{2 \left( m(x) + 4 \pi  P_x  x^3\right)}{x (x-2 m(x))} , \nonumber \\
&& \frac{d P_x}{d x} = - \frac{1}{G}  \frac{ d G}{d x} (\rho + P_x) + \frac{2}{x} (P_{\bot} - P_x ),
\label{eeq}
\end{eqnarray}
where energy density and pressures are defined by 
${\bf \widetilde{T}^{\mu}\,_{\nu} }=G^{\mu}\,_{ \nu}/8 \pi = {\rm Diag}(- \rho, P_x, P_{\bot}, P_{\bot})$
using  the coordinates ($t, x, \theta, \phi$) and the metric (\ref{metricmass}).
The first equation in (\ref{eeq}) shows that $m(x)$ is the mass inside a shell of radius $x$ and 
from the first equation in (\ref{gxx}) we can extract the function $m(x)$
($r=r(x)$),
\begin{eqnarray} \hspace{0.08cm}
m(x) = \frac{1}{2} \sqrt{\frac{a_o^2}{r^2}+r^2} \left(1-\frac{\left(r^4-a_o^2\right)^2 (r- r_-)
   (r- r_+)}{\left(a_o^2+r^4\right)^2 (r+r_*)^2}\right) \hspace{-0.08cm},
\nonumber 
\end{eqnarray}
where $r = r(x)$ is defined implicitly by the relation (\ref{x}).
The mass $m(x)$ above tends to the ADM mass for $r \rightarrow + \infty$ and 
to the dual ADM mass for $r \rightarrow 0$ (or $x \rightarrow + \infty$),
\begin{eqnarray}
m(x) \rightarrow  \left\{ \begin{array}{ll} 
         m(1+P)^2 &{\rm for} \,\,\, 
         x \rightarrow + \infty \, , \,\,\, r > \sqrt{a_o}, 
         \vspace{0.1cm}
         \\
                 \frac{a_o (1+P)^2}{4 m P^2} & 
         {\rm for} \,\,\,  x \rightarrow + \infty \, , \,\,\, r < \sqrt{a_o} . 
        \end{array} \right. 
\label{deltalimits}
\end{eqnarray}
The exact relation between the physical radial coordinate $x$ and $r$ is 
\begin{eqnarray}
r =  \left\{ \begin{array}{ll} \sqrt{\frac{x^2}{2}-\frac{\sqrt{x^4 - 4 \, a_o^2}}{2}  }
         & \,\,\, {\rm for} \,\,\,\,  r < \sqrt{a_o}, 
         \vspace{0.1cm}
         \\
                 \sqrt{\frac{x^2}{2} + \frac{\sqrt{x^4 - 4 \, a_o^2}}{2} }
         & \,\,\, {\rm for} \,\,\,\,  r > \sqrt{a_o} . 
        \end{array} \right. 
\label{rx}
\end{eqnarray}

\section{Vaidya 
collapse}
\label{dynamical}

We will proceed by combining the static metric with a radially ingoing null-dust,
such that we obtain a dynamical space-time for a black hole formed
from such dust. In the present model this process, usually described by the Vaidya metric \cite{Vai}, will
 have corrections, negligible in the
asymptotic region, but crucial to avoid the formation of a singularity in the strong-curvature
region. The metric constructed this way in the following is not a strict solution of
the LQC minusuperspace  equations in the vacuum. 
In other words,
keeping in mind what we said in the previous section, 
 it is a solution of the Einstein equations with an effective energy tensor which 
depends also on the time variation of the mass.

We use the radial coordinate $r$ of the previous section and we start by making a coordinate transformation and rewrite the static space-time in terms
of the ingoing null-coordinate $v$. It is defined by the relation $d v = d t + d r/\sqrt{F(r) G(r)}$,
which can be solved to obtain an explicit expression for $v$. The metric then takes the form
\begin{eqnarray}
 d s^2 = - G(r) d v^2 + 2 \sqrt{\frac{G(r)}{F(r)}} \, d r  dv + H(r) d \Omega^2 ~.
\end{eqnarray}
Now we allow the mass $m$ in the static solution to depend on the advanced time, $m \to m(v)$. Thereby, we
will assume the mass is zero before an initial value $v_a$ and that the mass stops increasing at $v_b$. We can then, as
before, use the Einstein equations $G=8\pi \widetilde{T}$ to obtain the
effective quantum gravitational stress-energy tensor $\widetilde{T}$. $\widetilde{T}^v_{\; v}$ and $\widetilde{T}^r_{\; r}$
do not change when $m(v)$ is no longer constant. The transverse pressure $\widetilde{T}^\theta_{\; \theta} = \widetilde{T}^\phi_{\;\phi}$
however has an additional term
\beqn
\widetilde{T}^\theta_{\;\theta}(m(v)) = \widetilde{T}^\theta_{\; \theta}(m) -  \frac{P r^2 m^{\prime}(v)}{2 \pi (r + 2 m(v) P)^4}~ ,
\label{transverse}
\eeqn
where $m'=dm/dv$. Because of the ingoing radiation, the stress-energy-tensor now also has an additional non-zero component, $\widetilde{T}^r_{\; v}$, which
describes radially ingoing energy flux
\begin{equation}
G^r_{\; v} = \frac{2 (1 + P)^2 r^4 (r^4 - a_o^2) (r - r_*(v) ) m^{\prime}(v)}{(a_o^2 + r^4)^2 (r + r_*(v))^3}~ .
\end{equation}
Notice that also in the dynamical case, trapping horizons still occur
where $g^{rr}=F(r,v)$ vanishes \cite{bhd,bhd2}, so we can continue to use the notation from
the static case just that $r_\pm(v)$ and $r_*(v)$ are now functions of $v$. 

This metric reduces to the Vaidya solutions at large radius, or for $\epsilon \to 0, a_o \to 0$.
However, in the usual Vaidya solutions, the ingoing
radiation creates a central singularity. But as we see here, with the quantum gravitational 
correction, the center remains regular.

We note that the ingoing energy flux has two zeros, one at $r=r_*(v)$ and one
at $r=\sqrt{a_o}$, and is negative between these.
What happens
is that the quantum gravitational correction works against the ingoing flux
by making a negative contribution until the effective flux has dropped to zero
at whatever is larger, the horizon's geometric mean $r_*$ or the location of the dual radius $r=\sqrt{a_o}$.
The flux then remains dominated by the quantum gravitational
effects, avoiding a collapse, until it has passed $r_*$ and the dual radius where it quickly approaches what looks
like an outgoing energy flux to the observer in the second asymptotic region.
%
%

\section{A scalar field on the LBH background}
\label{scalarfield}

The wave-equation for a massless scalar field in a general spherically symmetric curved space-time reads
\begin{eqnarray}
\frac{1}{\sqrt{ -g }} \partial_{\mu} \left( g^{\mu \nu} \sqrt{- g} \partial_{\nu} \Phi \right) =0,
\label{SF}
\end{eqnarray}
where $\Phi \equiv \Phi(r, \theta, \phi, t)$. Inserting the metric of the
self-dual black hole we obtain the following differential equation
\begin{eqnarray}
&& \hspace{-0.4cm} H(r) \left(2  \frac{\partial^2 \Phi }{\partial t^2}
-G(r) F'(r)  \Phi'   
\right) \label{eqscalar} \\
&& \hspace{-0.4cm} - 2 G(r) \left(   \frac{\partial^2 \Phi }{\partial  \theta^2}
+\cot  \theta   \frac{\partial \Phi }{\partial  \theta}
+\csc ^2 \theta  \frac{\partial^2 \Phi }{\partial  \phi^2}   \right) \nonumber \\
&& \hspace{-0.4cm} - F(r) \Big[ 
%
H(r) G'(r)  \Phi'   
+2   G(r) \Big( H'(r) \Phi'   
+ H(r)   \Phi''  \Big)  
\Big] = 0, \nonumber 
 \label{SFGFH}
\end{eqnarray}
where a dash indicates a partial derivative with respect to $r$.
Making use of spherical symmetry and time-translation invariance, we write the scalar field as
\begin{eqnarray}
\Phi(r, \theta, \phi, t) := T(t) \, \varphi(r) \, Y(\theta, \phi) \quad.
\label{dec1Phi}
\end{eqnarray}
omitting the indexes $l,m$ in the spherical harmonic functions $Y_{l m}(\theta, \phi)$.
The standard method of separation of variables allows us to split Eq.(\ref{SFGFH})
in three equations, one depending on the $r$ coordinate, one on the
$t$ coordinate and the remaining one depending
on the angular variables $\theta, \phi$,
%
\begin{eqnarray}
&& \hspace{-0.3cm}
\frac{\sqrt{G F}}{H} \frac{\partial}{\partial r}
\left( H \sqrt{G F} \,\,  \frac{ \partial \varphi(r)}{\partial r} \right)   \label{radial} 
= \left( G \frac{l(l+1)}{H}  - \omega^2 \right) \varphi(r),   \nonumber  \\
%
&&  \hspace{-0.3cm}
\hspace{0cm}  \left( \frac{\partial^2  }{\partial  \theta^2}
+\cot  \theta   \frac{\partial  }{\partial  \theta}
+\csc ^2 \theta  \, \frac{\partial^2  }{\partial  \phi^2}   \right) Y(\theta, \phi) 
\hspace{-0.0cm} = 
- K^2 Y(\theta, \phi), \nonumber \\
%
&& \hspace{-0.2cm} \frac{\partial^2 }{\partial t^2}  T(t) = - \omega^2 T(t),
\end{eqnarray}
where $K^2 = l(l+1)$.
To further simplify this expression we rewrite it by use of the tortoise coordinate $r^*$ implicitly defined by
\begin{eqnarray}
\frac{d r^*}{d r} := \frac{1}{\sqrt{GF}} ~.
\label{torto}
\end{eqnarray}
Integration yields the new radial tortoise coordinate
\begin{eqnarray}
&& \hspace{-0.5cm} r^* = r   - \frac{a_o^2}{r \, r_- r_+}
+ a_o^2 \frac{ \left( r_-  +  r_+ \right)}{ r_-^2
   r_+^2 } \log(r)  \\
&&   \hspace{-0.5cm}
- \frac{\left( a_o^2 + r_-^4\right)}{r_-^2 (r_+ - r_-)}   \log |r - r_-| 
+   \frac{\left(a_o^2 + r_+^4\right) }{r_+^2
   (r_+  -  r_-)} \log |r- r_+| ~. \nonumber 
\label{tortoise}
\end{eqnarray}
Further introducing the new radial field $\varphi(r) := \psi(r)/\sqrt{H}$
the radial equation (\ref{radial}) simplifies to
\begin{eqnarray}\label{simply}
&& \hspace{-0.5cm} \left[\frac{\partial^2}{\partial r^{* 2}} + \omega^2 - V(r(r^*)) \right] \psi(r) = 0,  \\
&& \hspace{-0.5cm}
V(r) =  \frac{ G K^2 }{H} 
+ \frac{1}{2} \sqrt{ \frac{G F}{H}} \left[ \frac{\partial}{\partial r} \left(  \sqrt{ \frac{G F}{H}}  \frac{ \partial H}{\partial r} \right) \right]. \nonumber
\end{eqnarray}
Inserting the metric of the self-dual black hole we finally obtain
\begin{eqnarray}
&& \hspace{-0.3cm}
V(r) = \frac{(r-r_-) (r-r_+)}{(r^4 + a_o^2)^4} \times \nonumber \\
&&\hspace{-0.3cm} \Big[r^2 \Big(a_o^4 \left(r
   \left( \, \left(K^2-2\right) r+r_- + r_+\right)+2 K^2 r r_*+K^2 r_*^2 \right) \nonumber \\
   &&  \hspace{-0.3cm}
   +2 a_o^2 r^4
   \Big(\left(K^2+5\right) r^2+2 K^2 r r_*+K^2 r_*^2-5 r (r_-+r_+) \nonumber \\
   && \hspace{-0.3cm}
   +5 r_- r_+\Big)+r^8
   \left(K^2 (r+r_*)^2+r (r_-+r_+)-2 r_- r_+\right)\Big)  \Big]. \nonumber
   \label{VS}
   \end{eqnarray}
The potential $V(r)$ is zero at $r = r_+$ and $r_-$ as for the classical Reissner-Nordstr\(\ddot{\text{o}}\)m black hole.
We therefore can follow the same analysis as for this case,  approximating $V(r(r^*))$ near the horizons via
\begin{eqnarray}
&& \hspace{-0.5cm} V(r^*) \propto e^{2 \kappa_+ r^*} \,\,\,\,\,  , \,\,\, {\rm for} \,\,\,  r \rightarrow r_+ \,\,\, {\rm or} \,\,\, r^* \rightarrow - \infty \, , \nonumber \\
&&  \hspace{-0.5cm} V(r^*) \propto e^{- 2 \kappa_- r^*} \, , \,\,\, {\rm for} \,\,\,  r \rightarrow r_- \,\,\, {\rm or} \,\,\, r^* \rightarrow + \infty , \nonumber \\
&& \hspace{-0.5cm} V(r^*) \rightarrow 0 \hspace{1cm}  , \,\,\, {\rm for} \,\,\, r \rightarrow 0 \,\, {\rm or} \,\, r \rightarrow + \infty.
\label{apprV}
\end{eqnarray}
A crucial quantity to study black hole evaporation is the surface gravity
\begin{eqnarray}
\kappa^2 = - \frac{1}{2} g^{\mu \nu} g_{\rho \sigma} \nabla_{\mu} \chi^{\rho} \nabla_{\nu}
\chi^{\sigma} 
= \frac{F \, (G'(r))^2 }{4 G} ,
\end{eqnarray}
where $'$ denotes the derivative respect to the radial coordinare $r$ and $\chi^{\mu}=(1,0,0,0)$ is a timelike Killing vector
in $r>r_+$ and $r<r_-$ but space-like
in $r_- <r < r_+$.
%
%
For the metric (\ref{g}) we  find the following values
\begin{eqnarray}
 \kappa_- =  \frac{4 m^3 P^4 (1-P^2)}{16 m^4 P^8 + a_o^2}, \,\,\,\,\,\,
 \kappa_+ = \frac{4 m^3 (1-P^2)}{16 m^4 + a_o^2},
\label{kpm}
\end{eqnarray}
for the surface gravity on the
inner and outer horizons.

\section{Particle Creation} \label{creation} 
We now work out the particle production in the background geometry describing 
the formation of a LBH.
The spacetime associated to the gravitational collapse to form a black hole is not everywhere
stationary and then we expect particle creation. The particle creation phenomenon 
is due to the non stationary gravitational collapse. 
However the spacetime will be stationary 
at late time and then particle creation is just a transient phenomenon
and we can forget all the details related to the gravitational collapse. 
We introduce the Hawking effect in the simplest possible scenario a la Vaidya 
explained in section (\ref{dynamical}).
The basic tools to evaluate the particle production are the Bogulibov transformations needed to connect the positive frequency modes of the field between the initial and final stationary regions .
The field can in fact be expanded in the initial stationary region as 
\begin{equation}
\Phi=\sum_{\omega} a^{\rm in}_{\omega} f_{\omega}+a^{\rm in\dagger}_{\omega} f^*_{\omega}
\label{phiin}
\end{equation}
 (the use of discrete modes from the continous one can be implemented smearing with suitable wave packets \cite {Haw1, Fabbri, wald})
 or in the final one as
 \begin{equation}
\Phi=\sum_{\omega} a^{\rm out}_{\omega} p_{\omega}+a^{\rm out\dagger}_{\omega} p^*_{\omega}
\label{phiout}
\end{equation}
where $a^{\rm in}_{\omega},a^{\rm out}_{\omega} $
 are the ladder operators verifying the usual commutation relations and $f_{\omega}$ and $p_{\omega}$ are the solutions of (\ref{simply}) in the initial and final regions respectively.

We define the {\em in} Fock space with the natural time $v$ at $I^-$.
The positive frequency modes which are solution of (\ref{simply}) at infinity ($I^-$) are 
\begin{eqnarray}
f_{\omega} (r, v) = \frac{e^{ - i \omega(r^* +  t)}}{4 \pi \sqrt{\omega} \sqrt{H} } = 
\frac{e^{ - i \omega v}}{4 \pi \sqrt{\omega} \sqrt{H} },
\label{modesmeno}
\end{eqnarray}
where $v = t + r^*$ and they obey the scalar product 
\begin{eqnarray}
&& \hspace{-1cm}  (f_{\omega} , f_{\omega'} ) = - (f_{\omega}^* , f_{\omega'}^* ) = \nonumber \\
&& \hspace{-1cm}- i \int_{I^-} 
dv \, H d \Omega (f_{\omega} \partial_v f_{\omega'}^* 
- f_{\omega'}^* \partial_v f_{\omega} ) = \delta(\omega -  \omega' ) 
\end{eqnarray}
and $(f_{\omega} , f_{\omega'}^* ) = 0$. We define also the $out$ Fock space at $I^+$ associated with the natural time parameter 
$u$. The positive frequency modes are
\begin{eqnarray}
p_{\omega} (r, u) = \frac{e^{ - i \omega(- r^* +  t)}}{4 \pi \sqrt{\omega} \sqrt{H} } = 
\frac{e^{ - i \omega u}}{4 \pi \sqrt{\omega} \sqrt{H} },
\label{pu}
\end{eqnarray}
where $u = t - r^*$ 
and the outgoing modes obey the normalization 
condition 
\begin{eqnarray}
&& \hspace{-1cm}(p_{\omega} , p_{\omega'} ) = - (p_{\omega}^* , p_{\omega'}^* ) = \nonumber \\
&& \hspace{-1cm}- i \int_{I^+} 
du H d \Omega (p_{\omega} \partial_{\bf u} p_{\omega'}^* 
- p_{\omega'}^* \partial_{\bf u} p_{\omega} ) = \delta(\omega - \omega' )
\end{eqnarray}
and $(p_{\omega} , p_{\omega'}^* ) = 0$. 
The modes $f_{\omega}$ and $p_{\omega}$ are solutions of (\ref{simply}) at $I^-$ and
$I^+$ respectively; we can also approximate $H \approx r^2$ in these
the asymptotic regions. 
We decided about $I^+$ as Cauchy surface but it is not properly correct.
We must include the future event horizon $H^+$ to have a complete Cauchy surface:
$I^+ \bigcup H^+$. The modes $p_{\omega}$ are not complete and we have to add those
that cross the future horizon $H^+$.
However we do not need   them to evaluate the particle production at $I^+$ because 
the result is insensitive to the  ingoing modes.
The next step is to calculate the Bogoliubov coefficients relating the ingoing and outgoing 
basis solutions $f_{\omega}$ and $p_{\omega}$. Assuming the frequency to be discrete
\begin{eqnarray}
p_{\omega} = \sum_{\omega'} A_{\omega \omega'} f_{\omega'} + B_{\omega \omega'} f_{\omega'}^*,
\end{eqnarray}
and
\begin{eqnarray}
A_{\omega \omega'} = (p_{\omega}, f_{\omega'} ) \,\,\,\,\, \& \,\,\, \,\,
B_{\omega \omega'} = - (p_{\omega}, f_{\omega'}^* ) 
\label{bogoLBH}
\end{eqnarray}
 which satisfy the following matrix relations again assuming discrete values for the frequency,
\begin{eqnarray}
&& A A^{\dagger} - B B^{\dagger} = 1, \nonumber \\
&& A B^{T} - B A^{ T} =0
\label{bogorel}
\end{eqnarray}
and the matrix elements are 
\begin{eqnarray}
A_{\omega \omega'} = - i \int_{I^-} dv H d \Omega (p_{\omega} \partial_v f^*_{\omega} - 
f^*_{\omega'} \partial_v p_{\omega}),  
\label{bogoA}
\end{eqnarray}
where, for  mathematical 
convenience we have chosen $I^-$  as the Cauchy surface to calculate the scalar
product that is insensitive to this choice.
The Bogulibov coefficients $A$ and $B$ can also be used to expand one of the two sets of creation and annihilation operators in terms of the other,
\begin{equation}
a^{\rm in}_{\omega}=\sum_{\omega'} A_{\omega'\omega} a^{\rm out}_{\omega'}+B^{*}_{\omega'\omega} a^{\rm out\dagger}_{\omega'}
\label{ain}
\end{equation} 
\begin{equation}
a^{out}_{\omega}=\sum_{\omega'} A^*_{\omega\omega'} a^{\rm in}_{\omega'}-B^{*}_{\omega\omega'} 
a^{\rm in\dagger}_{\omega'}
\label{aout}
\end{equation} 
If any of the $B_{\omega, \omega'}$ are non zero the particle content of the vacuum state at $I^-$
(which we indicate with the ket $| {\rm in} \rangle$) 
respect to the Fock space at $I^+$ is non trivial,
\begin{eqnarray}
\langle  {\rm in} | \hat{N}^{I^+}_{\omega} | {\rm in} \rangle = \sum_{\omega'} | B_{\omega, \omega'}|^2.
\label{number}
\end{eqnarray}
where $\hat{N}^{I^+}_{\omega}$ is the particle number operator at frequency $\omega$ at $I^+$.
In contrast, if all the coefficient $B_{\omega, \omega'}$ are equal to zero, the 
first of the relations (\ref{bogorel}) reduces to $A A^{\dagger} = 1$ and then the positive frequency mode
basis $f_{\omega}$ and $p_{\omega}$ are related by a unitary transformation and the annihilation operators (\ref{ain}) and (\ref{aout}) define the same vacuum.

\begin{figure}
 \includegraphics[width=8cm]{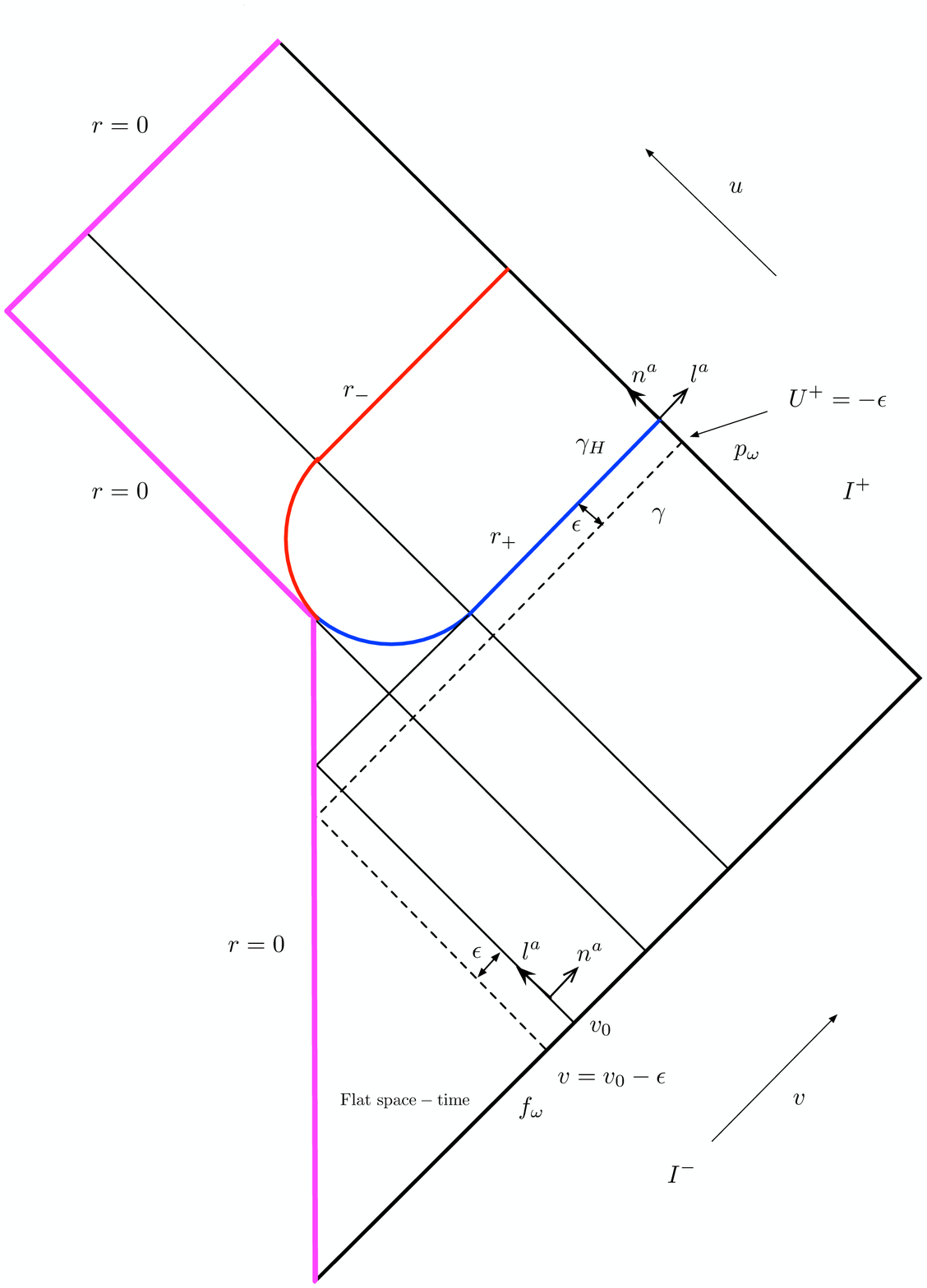} \hspace{-1cm}
\caption{Penrose diagram for the Vaidya collapse and slowly evaporation. The Penrose diagram for the
whole collapse and evaporation process will be given in the next figure.} 
\label{haw}
\end{figure}

To evaluate the products (\ref{bogoLBH}) we need to know the behavior of
the modes $p_{\omega}$ at $I^-$. For this propose we consider a {\em geometric optic 
approximation} in which the massless particle world-line is a null ray, $\gamma$, of constant 
$u$ and we trace this ray backwards in time from $I^+$ until $I^-$ (see Fig.\ref{haw}). 
The later it reaches $I^+$, the closer it must approach $H^+$.
The ray $\gamma$ is  one of the rays whose limit as $t \rightarrow + \infty$ is a null generator $\gamma_H$ of $H^+$.  To specify $\gamma$ we can then use  its affine distance from $\gamma_H$ along an ingoing null geodesic through $H^+$. This can be easly found using  the 
Kruscal-Type coordinates in the region outside the horizon $r_+ = 2 m$. These are defined by 
\cite{Modesto:2008im} 
\begin{eqnarray}
U^+ = - \frac{1}{\kappa_+} e^{- \kappa_+ u} \, \,\, , \,\,\,\,\,\,\,  
V^+ =   \frac{1}{\kappa_+} e^{\kappa_+  v}, 
\label{kruscal}
\end{eqnarray}
where $\kappa_+$ is the surface gravity calculated in (\ref{kpm}).
The affine parameter in the ingoing null geodesic closer to $H^+$ is $U^+  = - \epsilon$
and so using the first of (\ref{kruscal}) we find 
\begin{eqnarray}
u = - \frac{1}{\kappa_+} \log \epsilon + {\rm const}.
\end{eqnarray}
 substituting the previous expression in the solution  (\ref{pu}), we can see that it  oscillates rapidly at later times $t$ and this justifies the geometric optics approximation.
We need to match $p_{\omega}$ with the solution of the Klein-Gordon
equation near $I^-$. 
In the geometric optic approximation we just parallel-transport the vector $n^a$, tangent 
to the null geodetic which is ingoing at $H^+$, and $l^a$ which is the null generator of $H^+$,
back to $I^-$ along the continuation of $\gamma_H$.
We call $v_0$ the point where the continuation meets $I^-$ then the continuation 
of the ray $\gamma$, along the outgoing null geodesic, meets $I^-$ at $v = v_0 - \epsilon$ so 
\begin{eqnarray}
&& p_{\omega} = \frac{e^{ i \frac{\omega}{\kappa_+} \log(v_0-v) }}{4 \pi \sqrt{\omega} \, \sqrt{H} } \,\,\,\, {\rm for} \,\,\, v < v_0, \nonumber \\
&& p_{\omega} = 0 \hspace{1.9cm} \,\,\,\, {\rm for} \,\,\, v > v_0, 
\label{modespiu}
\end{eqnarray}
where $v_0$ is the latter time at which the field can reach infinity without entering in the black hole.
Next step is the calculation of the Bogolubov matrix A. Introducing (\ref{modespiu}) and (\ref{modesmeno})
in (\ref{bogoA}) 
we find 
\begin{eqnarray}
&& 
A_{\omega, \omega'} = \frac{(i \omega)^{-i \omega/\kappa_+}}{2 \pi \sqrt{\omega \omega'}} 
\Gamma \left(1+ i \frac{\omega}{\kappa_+} \right), \nonumber \\
&& 
B_{\omega, \omega'} = - i A_{\omega, - \omega'},
\end{eqnarray}
where we define $v_0 =0$. 
The Bogoliubov coefficient $A_{\omega, \omega'}$ is the fourier transform of 
a function that vanishes for $v > v_0$ than it is analytic in the lower half of the 
complex $\omega'$ plane. It has a logarithmic branch point in $\omega'=0$ 
then the branch cut extends into the upper half plane. Therefore, we have the following 
relation between the coefficients $A$ and $B$
\begin{eqnarray}
|A_{\omega, \omega'} | = e^{\frac{ \pi \omega}{\kappa_+}} | B_{\omega, \omega'}|.
\label{AeB}
\end{eqnarray}
Finally introducing (\ref{AeB}) in the first relation (\ref{bogorel}) we find 
\begin{eqnarray}
&& \hspace{-0.82cm} \delta_{\omega, \omega'} = (A A^{\dagger})_{\omega, \omega'} - (B B^{\dagger})_{\omega, \omega'} \nonumber \\
&& = \left[ e^{\pi(\omega + \omega')/\kappa_+} -1 \right]  (B B^{\dagger})_{\omega, \omega'} .
\label{radia}
\end{eqnarray}
Taking $\omega = \omega'$ 
the number of particles (\ref{number})
in the $\omega^{\rm th}$ mode at $I^+$ is
\begin{eqnarray}
\langle {\rm in} | \hat{N}^{I^+}_{\omega} | {\rm in} \rangle =  (B B^{\dagger})_{\omega, \omega}  = \frac{1}{e^{2 \pi \omega /\kappa_+} -1}.
\label{Planck}
\end{eqnarray}
The result (\ref{Planck}) coincides with the Planck distribution of thermal 
radiation for bosons at the temperature $T_{BH} = \kappa_+/2 \pi$.

\begin{figure}
\includegraphics[width=6.5cm]{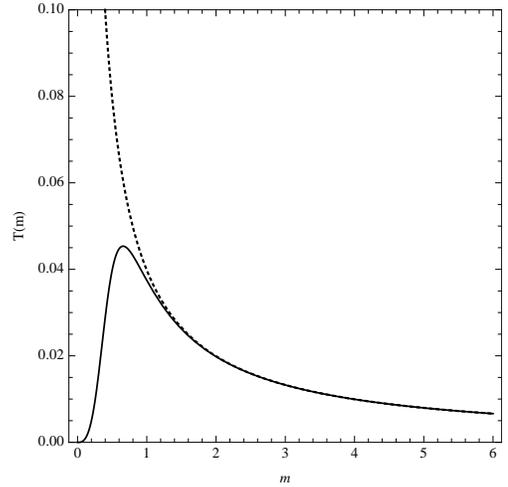}
\caption{Plot of the temperature versus the loop black hole mass. The dashed line represents the 
classical temperature $T=1/8 \pi m$.}
 \label{TempLBH}
\end{figure}

\paragraph*{Evaporation time.} 
\label{thermo}
The evaporation proceeds through the Hawking emission at $r_+$, and
the black hole's Bekenstein-Hawking temperature, given in terms of the surface gravity
$\kappa$ by $T_{BH}= \kappa/2 \pi$, yields \cite{Modesto:2009ve}
\begin{eqnarray}
T_{BH}(m) = \frac{(2m)^3 (1 - P^2)}{4 \pi [ (2 m)^4 + a_o^2]} ~.
\label{Temperatura}
\end{eqnarray}
This temperature coincides with the Hawking temperature in the limit of
 large masses but goes to zero for $m \rightarrow 0$ (Fig.\ref{TempLBH}).

The luminosity can be estimated by use of the
Stefan-Boltzmann law  $L(m)= \alpha A_H(m) T_{BH}^4(m)$,
where (for a single massless field with two degrees of freedom)
$\alpha = \pi^2/60$, and $A_H(m) = 4 \pi [ (2m)^2+ a_o^2/(2 m)^2]$ is the surface
area of the horizon. Inserting
the temperature, we obtain
\begin{eqnarray}
L(m) =
\frac{16 \, m^{10} \alpha \,(1-P^2)^4}{\pi^3
(a_o^2+16\, m^4)^3}~ .
\label{lumini}
\end{eqnarray}
The mass loss of the black hole is given by the following equation for the black hole mass 
\begin{eqnarray}
\frac{d m}{d v} = - L(m)
\label{flux}
\end{eqnarray}
and we can integrate its inverse to obtain
the mass function $m(v)$. The result of this integration with initial
condition $m(v = v_0) = m_0$ is
\begin{eqnarray}
&& \hspace{-0.8cm} \Delta v  =  \frac{(5 a_o^6 + 432 a_o^4 m^4 + 34560 a_o^2 m^8 - 
   61440 m^{12}) \pi^3}{720 m^9 (1 - P^2)^4 \alpha} \nonumber \\
 && \hspace{-0.8cm} -  \frac{(5 a_o^6 + 432 a_o^4 m_0^4 + 34560 a_o^2 m_0^8 - 
   61440 m_0^{12}) \pi^3}{720 m_0^9 (1 - P^2)^4 \alpha}~.
\label{v(m)}
\end{eqnarray}
In the limit $m \rightarrow 0$ this expression becomes $ \Delta v \approx a_o^6 \pi^3/(144 m^9 (1 - P^2)^4 \alpha)$,
and one thus concludes that the black hole needs an infinite amount of
time to completely evaporate. In the complete evaporation process, 
we neglected the backreaction for any value of the mass
%
because at $v \approx + \infty$, when $m \lesssim m_P$, $dm/dv << m$ and then, contrary to the classical case, such 
approximation is valid also in the final stages of evaporation.

\section{Collapse and Evaporation}
\label{evap}

We are now ready to combine the black hole formation and evaporation. As in section \ref{dynamical}, we divide space-time into
regions of advanced time. We start with empty space before $v_a$ then 
the mass increase from $v_a$ to $v_b$ since the gravitational collapse. 
For astrophysical black holes
this evaporation will proceed very slowly, 
and $m$ remains constant to good
accuracy at $m_0$, but at some later time, $v_c$, Hawking radiation becomes
relevant and $m$ decreases until it reaches zero again in an infinity time as we have seen in
the previous section. 

We thus have the partition $-\infty<v_a<v_b< v_c<\infty$ with
\begin{eqnarray}
\forall v\in(-\infty,v_a)&:&m(v)=0,\label{va} \nonumber \\
\forall v\in(v_a,v_b)&:& 
m'(v)>0, \label{vb}   \nonumber \\
\forall v\in(v_b,v_c)&:&m(v)=m_0,  \nonumber  \\
\forall v\in(v_c, + \infty)&:& 
m'(v)<0, \label{vd} \nonumber \\
{\rm for} \,\, v  \rightarrow + \infty
&:&m(v) \rightarrow 0.\label{ve}
\end{eqnarray}
The mass would immediately start to drop without incoming
energy flux and thus $v_a=v_b$, but stretching this region out will be more illuminating to
clearly depict the long time during which the hole is quasistable.

To describe the Hawking-radiation we will consider the creation of
(massless) particles on the horizon such that locally energy is conserved.
We then have an ingoing radiation with negative energy
balanced by outgoing radiation of positive energy. Both fluxes originate
at the horizon and have the same mass profile which is given by the Hawking
temperature. The area with ingoing negative density is again described
by an ingoing Vaidya solution, while the one with outgoing positive
density is described by an outgoing Vaidya solution.

The outgoing Vaidya solution has a mass-profile that depends on
the retarded time $u$ instead of $v$ and the mass decreases instead
of increases. The retarded time is defined by $du = d t - d r/\sqrt{F(r) G(r)}$. After
a coordinate transformation, the metric reads
\begin{equation}
ds^2= -G(r,u) du^2 - 2 \sqrt{\frac{G(r,u)}{F(r,u)}} du dr + H(r) d\Omega^2~,
\end{equation}
where $F(r,u)$ and $G(r,u)$ have the same form as in the static case (\ref{g}) 
 but with $m$ replaced by a function $m(u)$.
We fix the zero point of the retarded time $u$ so that $r=r_+$
corresponds to $u_c=v_c$. Then there is a static region with total
mass $m_0$ for $v>v_c$, $u<u_c$. Note that since the spacetime described here
has neither a singularity nor an event horizon, we can
consider pair creation to happen directly at the trapping horizon instead
of at a different timelike hypersurface outside the horizon,
as done in \cite{His}. We have in this way further partitioned spacetime in regions,
broken down by retarded time:
\begin{eqnarray}
\forall u<u_c&:&m(u)=m_0 ~, \nonumber \\
\forall u>u_c&:& 
m'(u) <0 ~.
\end{eqnarray}

\begin{figure}
\includegraphics[width=8.5cm]{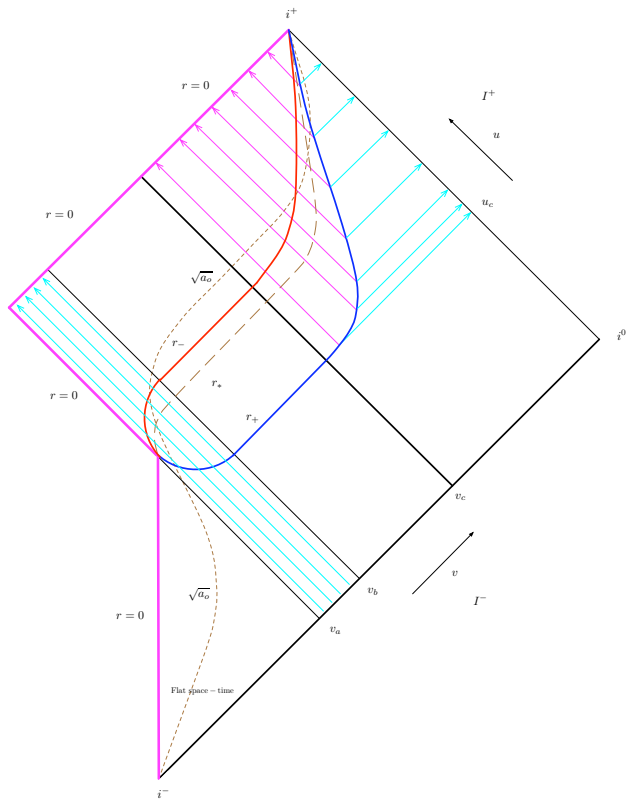}
\caption{Penrose diagram for the formation and evaporation of the regular black hole
metric. The red and dark blue solid lines depict the two trapping horizons $r_-$ and $r_+$. The
brown, dotted line is the curve of $r=\sqrt{a_o}$ and the brown, long dashed one is $r_*$.
The light blue arrows represent positive energy flux, the magenta arrows negative energy flux.} \label{bbh}
\end{figure}

Now that we have all parts together, let us explain the complete dynamics
as depicted in the resulting causal diagram Fig.\ref{bbh}.

In the region $v<v_a$ we have a flat and empty region, described
by a piece of Minkowski-space. For all times $v > v_a$, the inner and outer trapping
horizons are present. These horizons join
smoothly at $r=0$ in an infinite time and enclose a non-compact region of trapped surfaces. 

A black hole begins to form at $v=v_a$ from null dust which has collapsed completely at $v=v_b$ to a
static state with mass $m_0$. It begins to evaporate at $v=v_c$, and the complete
evaporation takes an infinite amount of time. The observer at ${I}^+$ sees particle
emission set in at some retarded time $u_c$. 
The region with $v>v_c$ is then
divided into a static region for $u<u_c$, and the dynamic Vaidya region for $u>u_c$, which
is further subdivided into an ingoing and an outgoing part.

As previously mentioned, the radially ingoing flux (light blue arrows) in the collapse region is
not positive everywhere due to the quantum gravitational contribution. It has
a flipped sign in the area between $r_*$ (black short dashed curve) and $r=\sqrt{a_o}$ (brown dotted
curve) which is grey shaded in the figure. Likewise, the ingoing negative flux during
evaporation (magenta arrows) has another such region with flipped sign. It is in this region,
between the two horizon's geometric mean value $r_*$ and the dual radius corresponding to the minimal area, that the quantum
gravitational corrections noticeably modify the classical and semi-classical
case, first by preventing
the formation of a singularity, and then by decreasing the black hole's temperature
towards zero in an infinity time.

\section{Unitarity Problem}
\label{unitarity problem}
In this section following Wald book \cite{wald}, we resume the main features of the loss of information and unitarity of classical black holes. 
The evolution of a physical system, represented by a quantum state $| {\rm in} \rangle$ in an initial Cauchy surface $\Sigma_i$ in Minkowski space, is given by a unitary operator that maps it into a final state $|f\rangle$ in a final Cauchy surface $\Sigma_f$.
Let's see what happens when a classical black hole is present:
The Cauchy surface $\Sigma_f$ can split into two surfaces: $\Sigma_f=\Sigma_{\rm int}\bigcup \Sigma_{\rm out}$ of which the first lies inside the black hole and the second lies outside. We can also think as limiting case $\Sigma_{\rm int}=\Sigma_{H^+}$ and $\Sigma_{\rm out}=\Sigma_{I^+}$. Now if $| {\rm in} \rangle=|0\rangle_{I^-}$ we know from \cite{wald} that we can formally write 
\begin{eqnarray}
U |0\rangle_{I^-} = \prod_{i}\left\{\sum^{\infty}_{N=0} e^{-\frac{N\pi\omega_i}{\kappa} } |N_i\rangle_{\rm int}\otimes |N_i\rangle_{\rm out} \right\} , 
\label{formal}
\end{eqnarray}
with $U$ given by a unitary transformation relating the vacuum of the initial Fock space to the $N$ particle states of the final Fock space. The form of the right hand side is  
due to the presence of a bifurcated Killing horizon and the consequent different notions of time on the initial and final Cauchy surfaces.
Note that the expression (\ref{formal}) is formal because the state on the right hand side fails to be normalizable if the condition 
\begin{equation}
{\rm Tr}( B^{\dagger} B) < \infty
\label{condition}
\end{equation}
is not fulfilled \cite{wald}. This condition implies that the total number of particles produced in the transition between the initial and final state is finite. If it is not satisfield the initial and final Fock spaces are unitary inequivalent, neverthless the expression  (\ref{formal}) can still be used to describe approximated states (see \cite{wald}). Note in fact that in both the usual Unruh and Hawking effects, this is the case.  

\begin{figure}
\includegraphics[width=8.5cm]{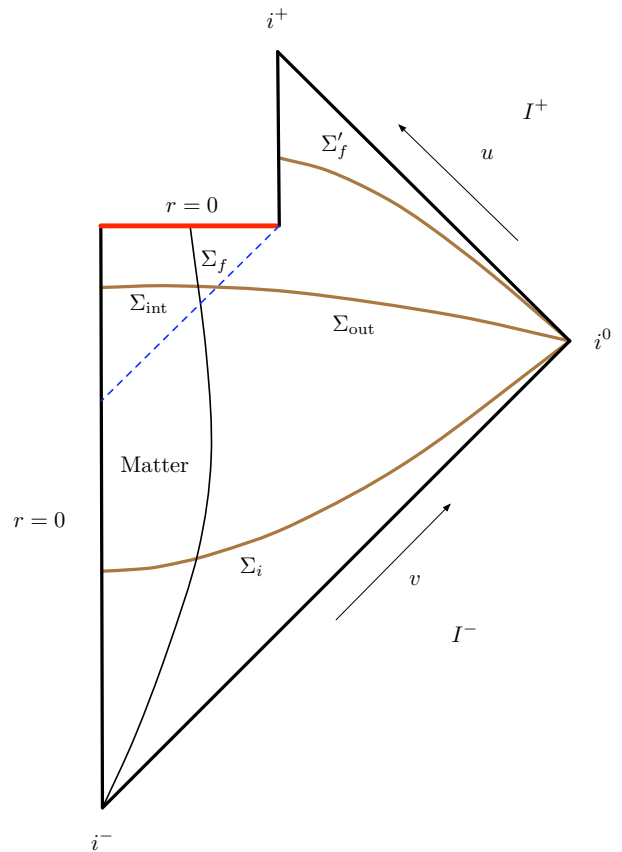}
\caption{Hawking scenario}
 \label{Hawkingscenario}
\end{figure}

Now the presence of the horizon obliges the external observer to \emph{trace} on the internal degrees of freedom, transforming the initial pure state $|{\rm  in} \rangle$ in a density matrix $\rho$. This is the information loss. Thus it is the horizon that causes the presence of a mixed state with an associated temperature, as a consequence of our ignorance of the complete system. The state that the observer at $I^+$ sees will not only be mixed but also made of uncorrelated radiation (stochastic thermal radiation).
The important point however is that the external observer loses the correlation between the two regions due to the tracing operation: \emph{but the correlations do exist!}. Altought the information is lost we can claim that the quantum unitary evolution is preserved at a fundamental level. If we were able to look inside the horizon we would see the correlations and the state would continue to be pure. 
The particles propagating to infinity are strongly correlated with particles that enter the black hole at early times. The presence of a density matrix 
to describe the state of the field in the exterior region from the algebraic point of view simply corresponds to the fact that the restricion of the state of the field  to the subalgebra associated to the $int \; D(\Sigma_{\rm out})$ (where $int$ means interior of a set and $D$ is the domain of dependence, see \cite{wald}) leads to a mixed state and this is  due to the fact that the "`domain of determinacy"' of $int \; D(\Sigma_{\rm out})$ is not the entire spacetime.
\emph{The  breakdown of the unitarity arises from the presence of a true singularity in the case of complete evaporation} see Fig.\ref{Hawkingscenario}. 

The disappearence of the black hole would completly remove from the spacetime the correlations (that could have preserved purity) hidden by the Horizon. As a consequence, on a surface $\Sigma'_f$  in the final spacetime, after the evaporation,  we would have really a mixed state also at a fundamental level. The conclusion is that not only we have information loss in the process $| {\rm in} \rangle\rightarrow \rho$, but also that this process \emph{is not Unitary}. The problem is that now at late time, the entire algebraic state of the field is mixed. In algebraic terms we can look at the whole process as an evolution of the state in a {\em time} labeled by the Cauchy surfaces. In the figure (\ref{Hawkingscenario}) we are looking at the evolution from the surface $\Sigma_f$ to the surface $\Sigma'_f$. Now the domain of determinacy of the hypersurface $\Sigma_f$ is the entire spacetime (see \cite{wald}). On the other hand $int \; D(\Sigma'_f)$ includes the entire future of the spacetime (after the evaporation), but its domain of determinacy  is not the entire spacetime, since it does not include the black hole region. The evolution corresponds then to the restriction of a pure state $\omega$ from the Weyl algebra of the entire spacetime to the subalgebra associated to the region $int \; D(\Sigma'_f)$ i.e evolution from a pure to a mixed state.
The entire problem can be summarized saying that we are evolving from a Cauchy surface to a surface that fails to be a Cauchy surface for the whole spacetime.
In the next section we will see in which sense these problems can be cured in our model.

\section{Unitarity Restored} \label{unitary}
We have shown in this paper that LBHs evaporate emitting particle in thermodynamical 
equilibrium at the temperature $T_{BH} = \kappa_+/2 \pi$. 
The black hole evaporation time is infinite
(\ref{v(m)}). 
As we have seen in the previous section, the black hole paradox presents essentialy two features:
the loss of information, strictly linked to the existence of an horizon, and the not unitary evolution, that is 
essentially related to the presence of a singularity.
The effective spacetime that we have described in the previous sections seems to be a good candidate to solve both problems.
The spacetime is in fact singularity free and the Bogoulibov coefficients evolve with the mass allowing a unitary tranformation between the initial and final vacuum.
Let's see in detail the behaviour of these coefficients: 
for small but non zero 
black hole mass we have thermal radiation 
but when the mass goes to zero, in an infinite amount of time, the 
unitarity is restored. More concretely looking at the 
surface gravity $\kappa_+$, we see the mass goes to zero (in an infinite time) 
together with the temperature. 
This is the crucial difference 
with the classical explosive case where the temperature goes to infinity when the mass 
reduces to zero.
From the Planck spectrum in (\ref{Planck}) we deduce that in an infinite amount of time
\begin{eqnarray}
\lim_{m \rightarrow 0}  (B B^{\dagger})_{\omega, \omega}  = 0 \,\,\,\,\,\, \forall \,\, \omega,
\end{eqnarray}
because $\kappa \rightarrow 0$ for $m\rightarrow 0$.
Since  $(B B^{\dagger})_{\omega, \omega}$ is positive semi-define, this vanish iff $B = 0$.
Let us now recall another useful relation between the initial and final Fock spaces. 
Using the Bogoliubov transformation between creation and annihilation operators respectively 
at $I^-$ and $I^+$, we can find the relation between the vacuum state $| {\rm in} \rangle$ (at $I^-$) 
and the vacuum state $| {\rm out} \rangle$ (at $I^+$), 
\begin{eqnarray}
| {\rm in } \rangle = \langle {\rm out} | {\rm in} \rangle \, 
e^{  \frac{1}{2} \hat{a}^{{\rm out} \dagger}_{\omega} V_{\omega \omega'} \hat{a}^{{\rm out} \dagger}_{\omega'}} \,
| {\rm out} \rangle,
\label{inout}
\end{eqnarray}
where the matrix $V$ is 
\begin{eqnarray}
V_{\omega \omega'} = - B^*_{\omega \omega''} A^{-1}_{\omega'' \omega'}.
\end{eqnarray}
Now for 
$m \rightarrow 0$, $B \rightarrow 0$ and then the initial state evolves in itself 
\begin{eqnarray}
| {\rm in } \rangle = | {\rm out} \rangle.
\end{eqnarray}
We conclude the {\em unitarity is restored} in an infinite amount of time. 
Since $B =0$ the positive frequency {\em mode basis $f_{\omega}$ and $p_{\omega}$ are related by 
the unitary transformation $A$} as immediate consequence of the first relation in (\ref{bogorel}).
Note also that due to this behaviour the $|{\rm out} \rangle$ state, unlike in the usual  treatment, can be related to the initial vacuum state by an exact unitary transformation because the necessary condition (\ref{condition}) is satisfied.

To test the radiation emitted is thermal, for non zero black hole mass, 
we can calculate the probabilities of emitting
different numbers of particles. For instance, we can calculate \cite{Fabbri}
\begin{eqnarray}
\langle {\rm in} | \hat{N}_{\omega}^{I^+} \hat{N}_{\omega}^{I^+} | {\rm in} \rangle 
= \frac{e^{ -2 \pi \omega/\kappa_+} (1 + e^{ -2 \pi \omega/\kappa_+})}{(1 - e^{ -2 \pi \omega/\kappa_+})^2},
\label{NN}
\end{eqnarray}
which agrees with a thermal distribution. In a similar 
way we can find all the higher moments that coincide with the thermal probability 
\begin{eqnarray}
P(N_{\omega} ) = (1 - e^{- 2 \pi \omega/\kappa_+}) \, e^{- 2 \pi N \omega/\kappa_+}.
\label{probemit}
\end{eqnarray}
to emit $N$ particles in the mode $\omega$. If we wait for an infinity amount of time
the mass goes to zero together with the thermal emission probability  
\begin{eqnarray}
\lim_{m \rightarrow 0} P(N_{\omega} \neq 0) = 0 \,\,\, , \,\,\,\, \lim_{m \rightarrow 0} P(N_{\omega} = 0) = 1
\end{eqnarray}
and the particle emission stops after an infinite amount of time. 
Using (\ref{probemit}) the expectation value of the {\em out} particle number operator reads 
\begin{eqnarray}
\langle {\rm in} | \hat{N}^{I^+}_{\omega} | {\rm in} \rangle = \sum_{N =0}^{+ \infty} N P(N_{\omega}).
\end{eqnarray}
The emission probability yet meets the normalization to one in the zero mass limit.

To check the thermal properties of the radiation produced by the black hole
we calculate 
$$\langle {\rm in} | \hat{N}_{\omega}^{I^+} \hat{N}_{\omega'}^{I^+} | {\rm in} \rangle = 
\langle {\rm in} | \hat{N}_{\omega}^{I^+}  | {\rm in} \rangle \, \langle {\rm in} | \hat{N}_{\omega'}^{I^+} | {\rm in} \rangle,$$
this result shows the absence of correlations between different modes as typical of the thermal radiation,
then the quantum state at $I^+$ is exactly described by 
a thermal {\em density matrix}
\begin{eqnarray}
\rho = \prod_{\omega} (1 - e^{- 2 \pi \omega/\kappa_+} ) \sum_{N =0}^{+ \infty} e^{- 2 \pi \omega N /\kappa_+} 
| N_{\omega} \rangle \langle N_{\omega} |,
\label{densitymatrix}
\end{eqnarray}
where $| N_{\omega} \rangle$ is the state at $I^+$ with $N$ particles in the mode $\omega$.
We see that any measure at $I^+$ is described by the density matrix $\rho$ and the von Newmann 
entropy $S_{\rm rad}= - {\rm Tr} \rho \log \rho$ associated to the frequency mode $\omega$ reads
\begin{eqnarray}
 S_{\rm rad}^{\omega}  
 = \frac{2 \pi \omega/\kappa_+}{e^{2 \pi \omega/\kappa_+} - 1} 
- \log\left(   1  -   e^{- 2\pi \omega/\kappa_+}   \right), 
\label{entropyrad}
\end{eqnarray}
and the total entropy is $S_{\rm rad} = \int_0^{+ \infty} d \omega S_{\rm rad}^{\omega} = \pi \kappa_+/6$.
When the balck hole mass goes to zero the density matrix reduces to 
\begin{eqnarray}
\lim_{m \rightarrow 0} \rho = \prod_{\omega} | 0_{\omega} \rangle \langle 0_{\omega} | \equiv 
| {\rm in} \rangle \langle {\rm  in} |,
\label{unitaruty}
\end{eqnarray}
and the state is pure. The process in its totality is 
\begin{eqnarray}
| {\rm in} \rangle \rightarrow  \rho (m >0) \rightarrow 
| {\rm in} \rangle \langle {\rm in} |,
\label{unitaruty2}
\end{eqnarray}
showing the initial pure state at $I^-$ evolves to itself at $I^+$.
The radiation at $I^+$ is no more thermal in the limit $m \rightarrow 0$ and
the final state is not a mixed state; the complete evolution is unitary. 
The entropy (\ref{entropyrad}) goes to zero.

During its evaporation the black hole 
is in thermal equilibrium with the radiation outside
for each finite value of the mass. When the black hole mass  
reduces to zero (in an infinity amount of time) 
the matter outside is no more in a thermal state but in a pure state. 
Thermal radiation is present outside the black hole only for finite values of the black 
hole mass but when $m \approx 0$ the mixed state collapses to a pure state.

\subsection{Role of the observer and correlations}
In the previous section we have restriced our attention to the surface $I^+$, now we condider also the previously neglected modes.
When the black hole mass is non zero the pre state $| {\rm in} \rangle$ could appear 
mixed but it is still a pure state since we have to add the modes crossing the event horizon 
$H^+$. 
A proper expansion of the field is then 
\begin{equation}
\Phi=\sum_{\omega} a^{\rm out}_{\omega} p_{\omega}+a^{\rm out\dagger}_{\omega} 
p^*_{\omega}+a^{\rm int}_{\omega} h_{\omega}+a^{\rm int \dagger}_{\omega} h^*_{\omega}
\label{phioutfull}
\end{equation}
where $a^{\rm int}_{\omega}$ are the ladder operators for the incoming particles at the future horizon and $h_{\omega}$ are the corresponding modes.
The modes $h_{\omega}$ are not uniquely defined because there is no natural time parameter on $H^+$. A simple expression for these modes \cite{wald2,Fabbri} can be obtained 
reversing the sign of $v_0-v$ and $\omega$ in (\ref{modespiu}). With this choice the additional Bogulibov coefficients G, E defined by
\begin{eqnarray}
h_{\omega} = \sum_{\omega'} G_{\omega \omega'} f_{\omega'} + E_{\omega \omega'} f_{\omega'}^*,
\end{eqnarray}
where 
\begin{eqnarray}
G_{\omega \omega'} = (h_{\omega}, f_{\omega'} ) \,\,\,\,\, \& \,\,\, \,\,
E_{\omega \omega'} = - (h_{\omega}, f_{\omega'}^* ) 
\label{bogoLBH2}
\end{eqnarray}
have simple relations \cite{Fabbri} with the coefficients $A$ and $B$,
\begin{eqnarray}
A_{\omega \omega'} = e^{2i\omega' v_o} G_{\omega \omega'}^*\quad 
B_{\omega \omega'} = e^{-2i\omega' v_o} E_{\omega \omega'}^* .
\label{bogoLBH3}
\end{eqnarray}
The annihilation operators for the modes entering the black hole region are then
\begin{equation}
a^{\rm int}_{\omega}=\sum_{\omega'} G^*_{\omega\omega'} a^{\rm in}_{\omega'}-E^{*}_{\omega\omega'} a^{\rm in\dagger}_{\omega'}.
\label{aint}
\end{equation} 
Because we have ignored this sector we have lost all the possible correlations 
with the quanta entering into the black hole. If we properly take into account 
those modes, the purity of the $| {\rm in} \rangle$ state is restored.
The state (\ref{inout}) is now
\begin{eqnarray}
| {\rm in } \rangle = \langle {\rm fin} | {\rm in} \rangle \, 
e^{  \frac{1}{2} \hat{a}^{{\rm fin} \dagger}_{i} V_{i j} \hat{a}^{{\rm fin} \dagger}_{j}} \,
| {\rm fin} \rangle,
\label{infin}
\end{eqnarray}
with the index $i,j$ variyng over all the modes on the final Cauchy surface ($int$, and $out$), and  $| {\rm fin} \rangle=|0\rangle_{int}\otimes|0\rangle_{out}$.
Following \cite{Fabbri} we find an explicit relation between $| {\rm in} \rangle $ 
and the $N$-particles states with frequency $\omega$ at $I^+$ and $H^+$,
respectively $| N_{\omega}^{I^+} \rangle$, $| N_{\omega}^{H^+} \rangle$,
\begin{eqnarray}
| {\rm in} \rangle =  \prod_{\omega} \sqrt{1 - e^{- \frac{2 \pi \omega}{\kappa_+}  }} \, \, 
\sum_{N = 0}^{+ \infty} e^{- \frac{\pi N \omega}{\kappa_+}} 
| N_{\omega}^{I^+} \rangle \otimes  | N_{\omega}^{H^+} \rangle.
\label{entangle}
\end{eqnarray}
This relation shows we have an independent emission in frequencies of quantum entangled 
states representing outgoing and ingoing radiation. Note that for $m \neq 0$ the thermal density matrix 
is recovered evaluating the expectation values of any operator $\hat{O}$ at $I^+$,
for example 
\begin{eqnarray}
&& \hspace{-0.2cm}  \langle {\rm in} | \hat{N}_{\omega}^{I^+} |{\rm in }\rangle = 
{\rm Tr} ( \rho \hat{N}^{I^+} ) \label{recoverrho}\\
&&\hspace{-0.2cm} 
= {\rm Tr} \left[ \prod_{\omega} ( 1 - e^{- \frac{2 \pi \omega}{\kappa_+}} ) 
\sum_{N = 0}^{+ \infty} e^{ - \frac{ 2 \pi \omega N }{\kappa_+}} 
| N_{\omega}^{I^+} \rangle \langle N_{\omega}^{I^+} | \hat{N}^{I^+} 
\right]. \nonumber 
\end{eqnarray}

\begin{figure}
\includegraphics[width=8.5cm]{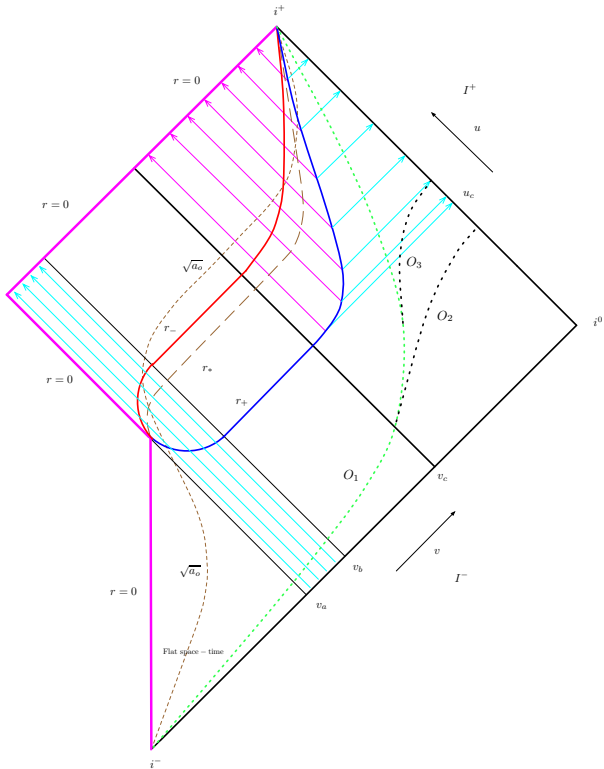}
\caption{Role of the observer and correlations}
 \label{osservatore}
\end{figure}
In Fig.\ref{bbh}  we can see that the evolution is unitary at a fundamental level, in fact the absence of singularity implies that any Cauchy surface of the spacetime is a Cauchy surface for the whole spacetime and the Hawking scenario of Fig.\ref{Hawkingscenario} is avoided. If we use the Cauchy surfaces $\Sigma$ as a "time label" for the spacetime describing the whole evaporation, we don't find any discontinuity in contrast to the Hawking scenario.
The presence of 
a continuum 
shrinking 
of the 
mass, allows us to describe the state (\ref{formal}) or equivalently (\ref{entangle}) in terms of a mass dependent splitting between internal and external spacetime of the final Cauchy surface  $\Sigma^m_f=\Sigma^m_{\rm int}\bigcup \Sigma^m_{\rm out}$:
\begin{eqnarray}
\hspace{-0.2cm}
U|0\rangle_{I^-}=\prod_{\omega}  \left\{ \sum^{\infty}_{N=0} e^{-\frac{N\pi\omega}{\kappa}} |N_\omega\rangle^m_{\rm int}\otimes |N_\omega\rangle^m_{\rm out} \right\} , 
\label{formal2}
\end{eqnarray}
which is based on the decoposition of the final one particle Hilbert space in terms of the direct sum of two one particle Hilbert spaces, one internal and the other external to the horizon, $H^m_{\rm fin}=H^m_{\rm int} \oplus H^m_{\rm out}$. When $m$ is zero the Hilbert space of solutions reduces to $H^0_{\rm fin}=H_{\rm r=0}\oplus H_{\rm I^+}$.
For 
$m \rightarrow 0$ the relation (\ref{formal2}) tends to 
\begin{equation}
U|0\rangle_{I^-}\rightarrow |0\rangle_{\rm int} \otimes  |0\rangle_{\rm out} 
\end{equation}
but also $B\rightarrow 0$ and as a consequence of (\ref{bogoLBH3} ) $E\rightarrow 0$.  In this limit then the initial vacuum reduces to 
\begin {equation}
|0\rangle_{in}=|0\rangle_{\rm out}\otimes|0\rangle_{\rm int}=|0\rangle_{\rm fin}
\end{equation}
and the density matrix to
the pure state (\ref{unitaruty}).

 We can better understand the properties of the radiation introducing an abserver in the classical region of the dynamical spacetime describing the collapse.
The situation is depicted in Figure (\ref{osservatore}). The Observer $O_2$ at $u<< u_c$ will not see any remarkable radiation because he is too far from the event horizon. 
Any Observer $O_3$ for $ - \infty > u  \geq u_c$ will see thermal radiation as an incoming flux of particles from the region where the matter collapsed. 
The essential point is the presence of an horizon:
 the pure state is perceived as a flux of particles because the observer is obliged to trace over the internal degrees of freedom, losing the information about the particles inside the black hole. The Observers $O_3$ will experience a thermal particle bath of increasing or decreasing temperature depending on the stage of the evaporation; the temperature is given by (\ref{Temperatura}). 
Finally for the observer $O_1$ (which follows a trajectory with constant radial coordinate) at $i^+$ 
the initial vacuum state evolves in itself as evident from (\ref{formal2}) when $\kappa \rightarrow 0$. 
The $O_1$ observer sees a gas of particles of increasing or decreasing temperature moving in time 
towards $i^+$.
During the evaporation the state 
is not pure because the trace on the
modes that cross the future horizon $H^+$ but when the mass goes to zero unitarity is restored.
It makes sense to speak about density matrix for any non zero value of the black hole 
mass but for $m \approx 0$ such operator reduces to a pure state in a natural way.
We wish to mention the fact that the observer in $i^+$ is in causal contact with both the radiation outcoming towards $I^+$ and the one ingoing to $r=0$, this fact in principle allows him to see the correlations between these modes and in this sense he could detect the previously hidden correlations needed to restore purity at all the times (a detailed study of of these correlations is currenlty under investigation \cite{to appear})
Note that any observer not located at $I^+$ would see a Temperature $T_{O}= \kappa/2 \pi\chi$ that differs from the temperature T by the ratio of "Killing time" to "proper time" for that observer, where $\chi$ is the redshift factor given by the square of the norm of the Killing vector.

We conclude this section with a comment.
We can compare the result in this section with what happens in the Rindler spacetime.
The Unruh temperature of the radiation is $T_U =g/2 \pi$ and it goes to zero when the
acceleration $g$ vanishes. The Bogoliubov transformation also goes to zero 
with the acceleration. For the LBH when the mass is smaller the Planck mass 
the behavior is the same, 
the temperature goes to zero with the decreasing black hole mass.
In other words the limits of zero acceleration and empty space coincide
leaving a pure state and the full process is unitary.

\section{Conclusions}

 We have calculated the particle creation by a loop black hole
 that is entirely singularity-free. 
 The analysis follows the classical one for the Schwarzschild spacetime 
 but with the crucial difference residing in the surface gravity of the black hole.
 The classical surface gravity (proportional to the temperature)
 diverges when the black hole mass goes to zero. For the LBH 
 instead it goes to zero with the mass but in an infinite time. 
 The approximation we used in this paper is that the matter fields
 (in particular we concentrated on the massless scalar field) 
 obey the usual wave equations with the Minkowski metric replaced 
 with a black hole spacetime metric $g_{\mu \nu}$ which is solution 
 of Einstein equations with an effective energy-tensor builds up a negative contribution that violates
the positive energy condition and prevents the formation of a singularity.
  We went trough the details of the Hawking calculation for a massless scalar field 
  that propagates in the given 
 fixed black hole background and we found the Bogoliubov 
 transformation as function of the new surface gravity.
 In an infinity amount of time the surface gravity goes to zero together with 
 the Bogoliubov transformation 
 and {\em the whole process, collapse and complete evaporation, is unitary}.
  The model we have studied in this paper is very simple and the analysis very similar 
 to the classical one but the quantum gravity properties of the metric imply that 
 the vacuum pure state at $I^-$ evolves in itself at $I^+$.
 We designed also 
the causal diagram for the complete process of collapse and evaporation. The value of
the scenario studied here is that it provides a concrete, calculable, model
for how quantum gravitational effects alter the black hole spacetime
and the particle flux to $I^+$.

\section*{Acknowledgements}
We wish to thank Thomas Thiemann and Alexander Stottmeister for many useful discussions.
E.Alesci would like to thank the Perimeter Institute for Theoretical
Physics, Waterloo, ON, Canada for the kind hospitality 
during a period of work on this project.
Research at Perimeter Institute is supported by the Government of Canada through Industry Canada
and by the Province of Ontario through the Ministry of Research \& Innovation.

\end{document}